\documentclass[12pt,preprint]{aastex}

\begin{document}

\title{$E=m c^2$ Without Relativity}

\author{Andrew Gould}
\affil{Department of Astronomy, The Ohio State University,
140 W.\ 18th Ave., Columbus, OH 43210}
\authoremail
%\email
{gould@astronomy.ohio-state.edu}

\singlespace

\begin{abstract}

The equivalence of mass and energy is indelibly linked with relativity,
both by scientists and in the popular mind.  Here
I prove that $E=m c^2$ by demanding momentum conservation of an object
that emits two equal electromagnetic wave packets in opposite directions
in its own frame.  In contrast to Einstein's derivation of this equation,
which applies energy conservation to a similar thought experiment,
the new derivation employs no effects that are greater than first order
in $v/c$ and therefore does not rely on results from Special Relativity.
In addition to momentum conservation, it uses only aberration
of starlight and the electromagnetic-wave momentum-energy relation
$p_\gamma= E_\gamma/c$, both of which were established by 1884.
In particular, no assumption is made about the constancy of the
speed of light, and the derivation proceeds equally well if one
assumes that light is governed by a Galilean transformation.
In view of this, it is somewhat puzzling that the equivalence
of mass and energy was not derived well before the advent of 
Special Relativity.  The new derivation is simpler and more transparent
than Einstein's and is therefore pedagogically useful.

\end{abstract}
\keywords{history and philosophy of astronomy -- relativity}
%\clearpage
%\newpage
 
\section{Introduction
\label{sec:intro}}

\citet{einstein05b} derived the equivalence of mass and energy ($E=m c^2$)
by considering an object of mass $m$ that simultaneously emits two 
electromagnetic packets,
each with energy $\Delta E/2$ in opposite ($\hat {\bf x}$ and $-\hat {\bf x}$)
directions.  By momentum conservation
in the rest frame of the object, it does not change its velocity after
emission.  Seen from a frame moving at velocity $-v\hat{\bf x}$ 
(i.e., along the axis
defined by the emissions), the two packets are each Doppler shifted
(in opposite directions), so that the total energy of these packets is
higher in the moving frame than the rest frame by $\Delta E(\gamma -1)$,
where $\gamma=(1-v^2/c^2)^{-1/2}$.  \citet{einstein05b} argued that
by energy conservation, the object must lose energy in the moving
frame by an amount that is greater than what it loses in the rest
frame by exactly this difference.
\citet{einstein05a} had already shown in his earlier paper 
introducing Special Relativity, that the kinetic energy of a mass $m$
moving at velocity $v$ is $E_k = m c^2(\gamma -1)$.  Appealing to this
result, \citet{einstein05b} concluded that the mass of the emitting 
object must decline from $m$ to $m' = m - \Delta m$, where 
$\Delta m = {\Delta E/ c^2}$.

Here I show that the same result can be derived from conservation
of momentum, without invoking any results from Special Relativity.
That is, the derivation uses only effects that are first order in
$v/c$, and does not employ the second-order effects that characterize
Special Relativity.

\section{Derivation of $E_0 = m_0c^2$
\label{sec:derivation}}

\begin{figure}
\plotone{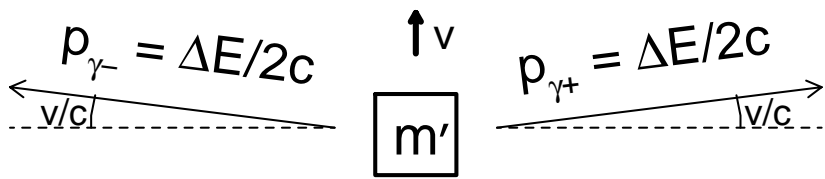}
\caption{\label{fig:frames}
Thought experiment proving $E=m c^2$ from momentum conservation
and without appealing to Special Relativity.
An object, originally of mass $m$, is pictured just after emitting two
electromagnetic wave packets, each of energy $\Delta E/2$.  Top panel:
by momentum conservation, the object remains at rest.  Bottom panel:
experiment viewed from frame moving downward at velocity $v$, in which
the packets have energy $\Delta E/2$ and so
momenta $p_{\gamma,\pm} =  \Delta E/2c$.
The object appears to be moving upward at $v$, both before and after
emission.  By aberration of starlight, the wave packets now travel
slightly upward at an angle $\theta=v/c$, and so have vertical
components to their momenta $(p_{\gamma,\pm})_z = \Delta Ev/2c^2$.
 From momentum conservation in the $z$ direction,
$m v = m'v + (p_{\gamma,+})_z + (p_{\gamma,-})_z = (m' + \Delta E/c^2)v$.
That is, the object has lost a mass $\Delta m = m - m' = \Delta E/c^2$.
}\end{figure}

Consider as above an object emitting the two 
electromagnetic packets that, viewed in its rest frame are equal
and opposite.  By momentum conservation, the dual ejection leaves 
the object at rest in this frame.  See Figure \ref{fig:frames}.
Now consider the same event from a frame that is moving
{\it perpendicular} (with velocity $-v\hat{\bf z}$) 
relative to the emission directions.  By symmetry, the wave
packets are still equal, but they are no longer opposite:
because of the aberration of starlight (first discovered by James Bradley in
1729), the packets will both appear to be moving slightly upward, at
an angle $\theta = v/c$.  Denote the emitted energies
of the packets {\it in the moving frame} by $\Delta E/2$, and denote the
mass of object in this frame before and after emission by $m$ 
and $m'$.

By a variety of arguments elaborated below, the magnitude of the 
momenta of the two packets in this frame are
\begin{equation}
p_{\gamma,\pm} = {\Delta E\over 2 c}.
\label{eqn:pgamma}
\end{equation}
Hence, because of aberration of starlight, the vertical components 
of these momenta will be (to first order in $v/c$)
\begin{equation}
(p_{\gamma,\pm})_z =
{\Delta E\over 2 c}\,{v\over c}.
\label{eqn:pgammamov}
\end{equation}
Equating the total $z$-momentum in the moving frame before and after emission 
yields,
\begin{equation}
m v =  m' v +  (p_{\gamma,+})_z +  (p_{\gamma,-})_z = 
\biggl(m' + {\Delta E\over c^2}\biggr)v,
\label{eqn:momcon}
\end{equation}
which can be solved to obtain,
\begin{equation}
\Delta m = m-m' = {\Delta E\over c^2}.
\label{eqn:emc2}
\end{equation}

Note that in carrying out this derivation, I explicitly ignored terms
higher than first order in $(v/c)$, in particular when I adopted
$p_z = p(v/c)$.  Hence, the result strictly applies only in the limit
$v\rightarrow 0$, i.e., in the rest frame.  This can be expressed as
an equivalence between energy and rest-mass,
\begin{equation}
E_0 = m_0c^2.
\label{eqn:em0c2}
\end{equation}
I address the question of how this result can be generalized to moving bodies
in \S~\ref{sec:moving}.

\subsection{Energy and Momentum of Light Packets
\label{sec:energymom}}

In the derivation, I used the relation between the energy $E_\gamma$
and momentum $p_\gamma$ for (monodirectional) electromagnetic fields,
\begin{equation}
p_\gamma = {E_\gamma\over c}.
\label{eqn:pegamma}
\end{equation}
Of course, this can be derived from Special Relativity, but the orientation
here is to derive equation (\ref{eqn:emc2}) with no recourse to Relativity,
nor to concepts of a similar vintage, such as photons.

\citet{jackson} recapitulates \citet{poynting}'s manipulations of 
Maxwell's equations to derive
the electromagnetic energy flux density ${\bf S} = (c/4\pi){\bf E \times H}$,
where ${\bf E}$ and ${\bf H}$ are the electric and magnetic fields.  He then
develops a similar manipulation of Maxwell's equations (together
with the Lorentz force law) to derive the momentum density 
${\bf g} = {\bf E \times B}/4\pi c$, where ${\bf B}$ 
is the magnetic induction.  Combining these two equations
for monodirectional electromagnetic waves in free space
yields equation (\ref{eqn:pegamma}).
This shows that this relation rests directly on the Maxwell/Lorentz
equations, although whether anyone actually derived the expression for
${\bf g}$ prior to the simplification of vector notation is not clear.

However, \citet{boltzmann} already uses $P=u/3$ for isotropic electromagnetic
radiation in his thermodynamic derivation of Stefan's law.  Here $P$ is the
pressure and $u$ is the energy density.  This expression already implies
$p=E/c$ for monodirectional electromagnetic waves.

\section{Generalization to $E=m c^2$
\label{sec:moving}}

As emphasized in \S~\ref{sec:derivation}, by carrying out the
derivation only to first order in $(v/c)$,  I ultimately restricted
its validity to bodies at rest.  Put differently, if the true relation
between mass and energy had the form, $E=m c^2(1 + \kappa(v/c)^2 + \ldots)$,
the derivation would have proceeded exactly the same way.  There
are two paths to generalizing the result to moving bodies.

The first is to adopt the results of Special Relativity.  This is
the approach of \citet{brown}, who derived $E=m c^2$ using
momentum conservation when light is emitted in an arbitrary direction.
In Special Relativity, equation (\ref{eqn:pgammamov}) is exact,
so the derived relation between mass and energy is exact to all
orders in $(v/c)$.  This approach is pedagogically useful: like
Einstein's derivation, it makes use of Special Relativity, but
it is simpler and more direct.

However, as a historical and logical exercise, one may also ask
how equation (\ref{eqn:em0c2}) could have been generalized if it
had been discovered prior to Special Relativity.  Such a generalization
follows from a simple thought experiment.  Imagine a box filled with
warm gas, whose thermal energy ultimately resides in the kinetic energy
of the atoms.  At the time, this picture was controversial but at
least some physicists (e.g., Boltzmann) held to it.  Light is emitted
from two holes in the box, similarly to the situation in 
\S~\ref{sec:derivation}.  The energy of the light packets is drawn
from the kinetic energy of the atoms in the box, some of which now
move more slowly.  By equation (\ref{eqn:emc2}), the box has lost
not only energy, but also mass.  However, since the box contains
no inter-atom potential energy, the mass (i.e., inertia) of the box must be the
sum of the mass (inertia) of the atoms in it.  As the number of these has not
changed, the mass of some of the atoms must have been reduced by 
exactly the amount of reduced mass of the box, which is exactly the
same as the kinetic energy lost from these atoms divided by $c^2$.
That is, kinetic energy also contributes to inertia.

\subsection{Derivation of Relativity From $E=m c^2$
\label{sec:reverse}}

Up to this point, I have derived $E=m c^2$ without ever making use
of \citet{einstein05a}'s postulate that $c$ is the same in all frames of
reference, nor of any of the results that he derived from this postulate.
I now show that Special Relativity, including the universality of $c$,
can be derived from this equation.

First, \citet{feynman63} shows that $E=m c^2$ leads to the growth
of inertia with velocity, $m= m_0\gamma$.  To permit clarification
of a subtle point, I repeat that derivation here,
beginning with the Newtonian equation
relating force to the increase of kinetic energy, $F=dE/dx$.
Using the definitions, $F=dp/dt$, $p=mv$, $v=dx/dt$, this can be written
$dE/dx = d(mv)/dt$, or
\begin{equation}
{dE\over dt} = v^2{d m\over d t} + m v{dv\over dt}.
\label{eqn:dedt}
\end{equation}
Substituting in the just derived $E=m c^2$ yields
\begin{equation}
{d(m c^2)\over dt} - {d m\over d t}v^2 = {m\over 2}\, {dv^2\over dt}.
\label{eqn:dmdt}
\end{equation}
At this point, there may be some question as to whether the
one may pull ``$c$'' out of the derivative, since it has not yet
been shown to be ``constant''.  But $c$ is a {\it constant} in any one
frame: the point that has not yet been addressed is whether it
is {\it invariant} under frame changes.  In the present case, the
observer is not changing frames: it is the mass that is accelerating.
The quantities $E$, $m$, $v$,  and $c$ are all as measured in the
observer frame, which is inertial.  We then obtain,
\begin{equation}
{d m\over m} = {d (v^2)\over 2(c^2 - v^2)},
\label{eqn:dmdt2}
\end{equation}
whose solution is
\begin{equation}
m = {m_0\over \sqrt{1-(v/c)^2}},
\label{eqn:mm0}
\end{equation}
where $m_0$ is an integration constant, which we identify with the
rest mass.

From this point, it is straightforward to derive the other relations
of Special Relativity by well-known arguments.  For example, as
a fast train passes by, a passenger and a bystander each throw tennis
balls transverse to the motion of the train (with equal strength)
in such a way that they hit and each bounces back directly to its 
respective thrower.  The balls must each return at the speed they 
were launched or the train passenger could detect her own motion.
Thus, they must have equal and opposite momenta.  The bystander
reckons that the passenger's ball is more massive and therefore
concludes it's transverse velocity is smaller, which can only
be true if time passes more closely.  By similar traditional arguments,
one can go on to derive length contraction, etc.  In this way,
one can prove that the speed of light is the same in all frames
of reference rather than assuming it.

\section{Discussion
\label{sec:discuss}}

While the result obtained here is obviously not new, there are three
reasons for establishing this result using a new derivation.  First,
the expression $E=m c^2$ is zeroth order in $v/c$, in sharp contrast
to the majority of results from Special Relativity, which are second
order.  It seems more elegant, therefore, to derive this expression
using first-order arguments, rather than relying on second-order
expressions.  Second, because the derivation is more elegant, 
it has pedagogical value, i.e., it is easier to transmit to students.
Third, because the derivation is independent of Special Relativity,
it raises the question of why $E=m c^2$ was not derived earlier than
1905.  In particular, the elements needed to derive it 
(momentum conservation, aberration of starlight, and the proportionality
between electromagnetic energy and momentum) were all in place
by 1884.  Indeed, once one realizes that electromagnetic waves have
momentum (even if one does not yet know the exact expression for
this quantity), it follows immediately from momentum conservation
and aberration of starlight that a light-emitting object must lose mass.

As reviewed by \citet{pais}, during the 25 years before Special Relativity
there were many efforts to express the mass of particles in terms
of their energy divided by $c^2$.  But these differed from the arguments
given here (and that I have argued could have been given at least as
early as 1884) by two important features.  First, they generally
centered around evaluations of the ultimately rather nebulous 
electromagnetic self-energy of charged particles rather than the 
kinetic properties of all matter (charged or neutral).  Second, these
evaluations did not recognize (at least explicitly) that when an
object emitted energy, it also lost mass.  Indeed, the very complexity
of the arguments developed in this era compared to the absolute
simplicity of the derivation in \S~\ref{sec:derivation} makes it even
more puzzling why no one hit on the latter.

%\begin{equation}
%\label{eqn:}
%\end{equation}

\acknowledgments
I thank Julio Chanam\'e, Subo Dong, and David Weinberg for valuable 
discussions.
This work was supported by grant AST 02-01266 from the NSF.

\bigskip
%\clearpage

\clearpage

\end{document}